\documentclass[intlimits,twoside,a4paper]{article}

\usepackage{graphicx}

\usepackage[cp1251]{inputenc}

\usepackage[eqsecnum]{cmpj3}

\issue{2017}{20}{2}{23302}
\doinumber{10.5488/CMP.20.23302}

\title[Influence of C$_{6}$H$_{4}$(OH)$_{2}$ isomers on water disinfection]%
{Influence of C$_{6}$H$_{4}$(OH)$_{2}$ isomers on water disinfection by photocatalysis: a computational study}
\author[K. Elmenaouar, R. Benbrik, A. Aamouche]{K. Elmenaouar, R. Benbrik, A. Aamouche\thanks{Corresponding author, E-mail: a.aamouche@uca.ma.} }
\address{MSISM Research Team, Physics Department, Polydisciplinary Faculty of Safi, Cadi Ayyad University, \\Sidi Bouzid, 46000, Safi, Morocco   }

\date{Received January 18, 2017, in final form May 8, 2017}

\begin{document}

\maketitle
\begin{abstract}
Solar disinfection by photocatalysis is one of the promising methods used for drinking water disinfection. It leads to the destruction of bacteria like {\it Escherichia Coli} ({\it E.~Coli}). In this paper, we compare our theoretical results with experimental ones done previously by A.G. Rinc{\'o}n and his colleagues concerning the order of decay of C$_{6}$H$_{4}$(OH)$_{2}$ isomers in the presence of titanium dioxide TiO$_{2}$, and show the influence of optical properties of those molecules on {\it E.~Coli} inactivation. According to the adsorption energy parameter, we find that catechol has the highest adsorption degree on titanium dioxide, followed by resorcinol, and finally hydroquinone. Three dihydroxybenzene isomers absorb photons belonging to ultraviolet (UV) range. The lowest absorption energies of resorcinol, catechol and hydroquinone are respectively 3.42, 4.44 and 4.49~eV.
\pacs 31.15.Ew, 33.80.b, 82.20.Wt, 78.20.Ci, 82.45.Jn, 82.50.Hp
\end{abstract}

\section{Introduction}
Water plays a crucial role in human life. Hence, for many decades, different studies have focused on its treatment by controlling the microbiological and chemical substances that can affect the safety of this essential element for life and human health using different techniques. Nevertheless, in many countries, environmental reasons and the cost of chlorination have discouraged the use of some conventional methods. Consequently, some alternative methods of disinfection \cite{Coo93,Ric96}, and new modifications in the conventional treatment have been proposed.

Solar photocatalytic process for water and wastewater treatment has emerged as a potential technology leading to total destruction of most bacterial population. In such process, titanium dioxide TiO$_2$ is found to be the most suitable catalyst thanks to its high efficiency, availability, low-toxicity, low cost and relatively high chemical stability. When photocatalyst titanium dioxide absorbs radiations with an energy higher than its band-gap, from sunlight or illuminated light source, electrons of the valence band become excited. The excess energy of the excited electrons promote the electrons to the conduction band, leaving positive holes ($h^+$). The formed positive-holes break apart the water molecules to form hydrogen gas and hydroxyl radicals OH$^{\bullet}$ which are highly toxic towards microorganisms \cite{Par12,Sch14,Sas10}.

In photocatalytic processes, titanium dioxide could be used in the form of powder, grains or fixed on a support in different geometries. To enhance the photocatalytic performance of TiO$_2$, a plethora of novel morphologies of TiO$_2$, such as nanosheets, nanotubes and nanowires have become increasingly synthetically controllable and can be designed to an unprecedented degree \cite{Hoa12,Lee14}. Synthesis strategies and key advantages of 1D nanostructures have been described in numerous excellent reviews \cite{Hab13, Sch14, Roy11, Cha08, Li15, Tan15}. Thereby, in our present work, we focus on one of the most investigated morphologies over recent years:  [010] {\it nanowire}, constructed by the most stable surface (101) \cite{Laz01}, and the most reactive one (001). The latter surface gives rise to a better adsorption of molecules.

In water systems, especially wastewaters, C$_{6}$H$_{4}$(OH)$_{2}$ isomers (or dihydroxybenzenes) exist due to a variety of natural (degradation products of the humicacids) and industrial sources \cite{Mil98}, where {\it E.~Coli} bacteria is found to be a common biological indicator of disinfection efficiency. Hence, in order to understand the influence of those isomers (catechol, hydroquinone and resorcinol) which could be introduced through a variety of natural (degradation products of the humicacids) and industrial sources \cite{Mil98} on the photocatalytic inactivation of this bacteria, we perform a computational study based on density functional theory (DFT) \cite{Hoh64} and time dependent density functional theory (TDDFT) \cite{Run84} of the structural and optical properties of those isomers. We explain the relation between the adsorption degree of each C$_{6}$H$_{4}$(OH)$_{2}$ isomer and its concentration decay. We also illustrate the effect of the optical properties of dihydroxybenzene isomers on {\it E.~Coli} inactivation by sunlight. The remainder of this paper is organized as follows. In section~\ref{ua-part}, we give details about the employed computational methodology. Structures of C$_{6}$H$_{4}$(OH)$_{2}$ isomers, free and adsorbed on TiO$_2$ nanowire are dealt with in section~\ref{ub-part}. In subsection~\ref{uc-part}, we studied the adsorption degree of each C$_{6}$H$_{4}$(OH)$_{2}$ isomer on titanium dioxide. The influence of optical properties of dihydroxybenzenes on {\it E.~Coli} inactivation is the subject of subsection~\ref{ud-part}.

%%%%%
\section{Computational details} \label{ua-part}
The calculations presented in this paper have been performed using density functional theory (DFT) and time-dependent density functional theory (TDDFT) in a plane-wave (PW) basis set and pseudopotential framework, using quantum ESPRESSO suite of programs \cite{Gia09,url1}. The Kohn-Sham (KS) orbitals are represented using basis sets consisting of PWs up to a kinetic energy cutoff of 30~Rydberg (Ry) and the charge density up to 300~Ry. The interaction between the nuclei and the electrons is modelled using Vanderbilt pseudopotentials named {\it X.pbe-van\_ak.UPF} (X=C, H and O) and {\it Ti.pbe-sp-van\_ak.UPF} which are available in quantum ESPRESSO website \cite{url1}. The electronic configurations taken into account explicitly in the calculations by making use of these pseudopotentials for carbon~(C), hydrogen~(H), oxygen~(O) and titanium~(Ti) atoms are $2s^22p^2$, $1s^1$, $2s^22p^6$ and $3s^23p^64s^23d^2$, respectively. The exchange-correlation functional is approximated using generalized gradient approximation (GGA) in the parameterization of Perdew, Burk and Ernzerhof (PBE) \cite{Per96}.

The adsorption procedure has been performed in three steps. Firstly, we relax the nanowire (O and Ti atoms of the nanowire) in an orthorhombic supercell of dimensions $a=20.109$~\AA, $b=28.153$~\AA, and $c=6.820$~\AA. Secondly, we fix the molecule, previously relaxed in the same supercell, on the surface of the nanowire, then we relax the molecule while keeping the nanowire fixed. Finally, the entire structure (nanowire+molecule) is relaxed, including the hydrogen atoms of hydroxyl groups of the molecule, placed on the nanowire. The relaxation has been performed using a grid of $2\times2\times4$ k-points.

 To calculate the optical properties of dihydroxybenzenes, we solve the linear response TDDFT equations using a recursive method \cite{Wal06,Roc08,Bar10,Wal08,Gho10}. We use the atomic positions of the relaxed molecules. The Brillouin zone was sampled with gamma point only, since k-point algorithm is not yet tested  in the corresponding program, turbo-TDDFT \cite{Mal11}, which is a part of the Quantum espresso package. The same DFT exchange-correlation functionals have been used in the adiabatic approximation (AXCA).

%%%%%%%%%%%%%%%%%%%%%%%%%%%%%%%%%%%%%%%%%%%%%%5%%%%%%%%%%%%%%%%%%%%
\section{Model system } \label{ub-part}
\begin{figure}[!t]
\centerline{\includegraphics[width=0.95\textwidth]{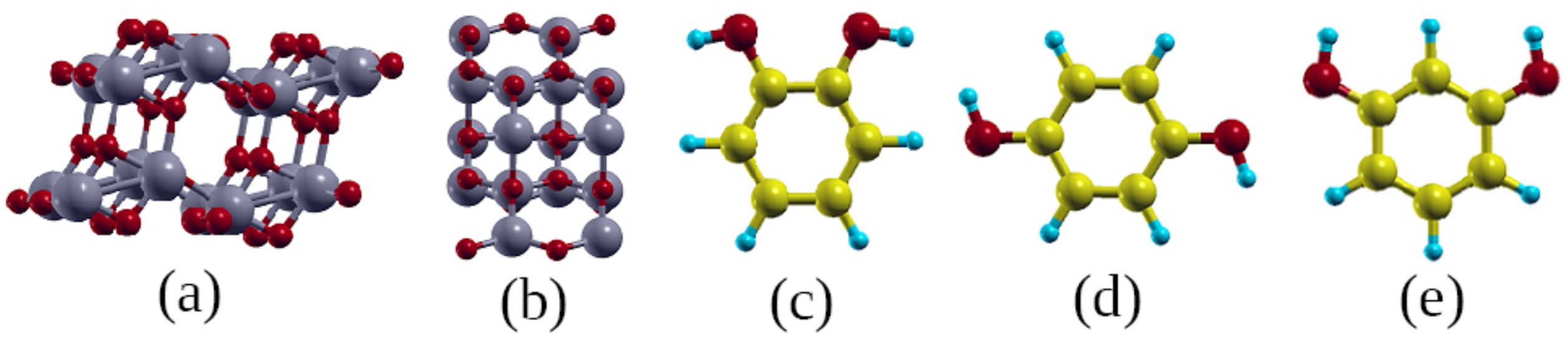}}
\caption{(Color online) Panel~(a) shows the cross section of the relaxed [010] anatase nanowire, while (b) shows its longitudinal section. Panels~(c), (d) and (e) represent the catechol, the hydroquinone and the resorcinol molecules, respectively, at the equilibrium. Blue, red, yellow and gray atoms correspond to the hydrogen, the oxygen, the carbon and the titanium atoms, respectively.} \label{fig-smp1}
\end{figure}

  For photocatalysis, we use the anatase phase of TiO$_2$, to construct TiO$_2$ nanowire. This phase is a dominant one according to the experimental results, thanks to its activity in light absorption. We model the TiO$_2$ nanowire with a segment comprising 16~TiO$_2$ units (48~atoms). This segment is repeated periodically in the [010] direction [panel (a) and (b) of figure~\ref{fig-smp1}]. In order to ensure negligible interactions with the periodic images of the wire, we use periodic boundary conditions in the direction perpendicular to the wire axis. The nanowire is separated from its periodic images by  7~{\AA} of vacuum.
%%%%%

\begin{figure}[!h]
\centerline{\includegraphics[width=0.75\textwidth]{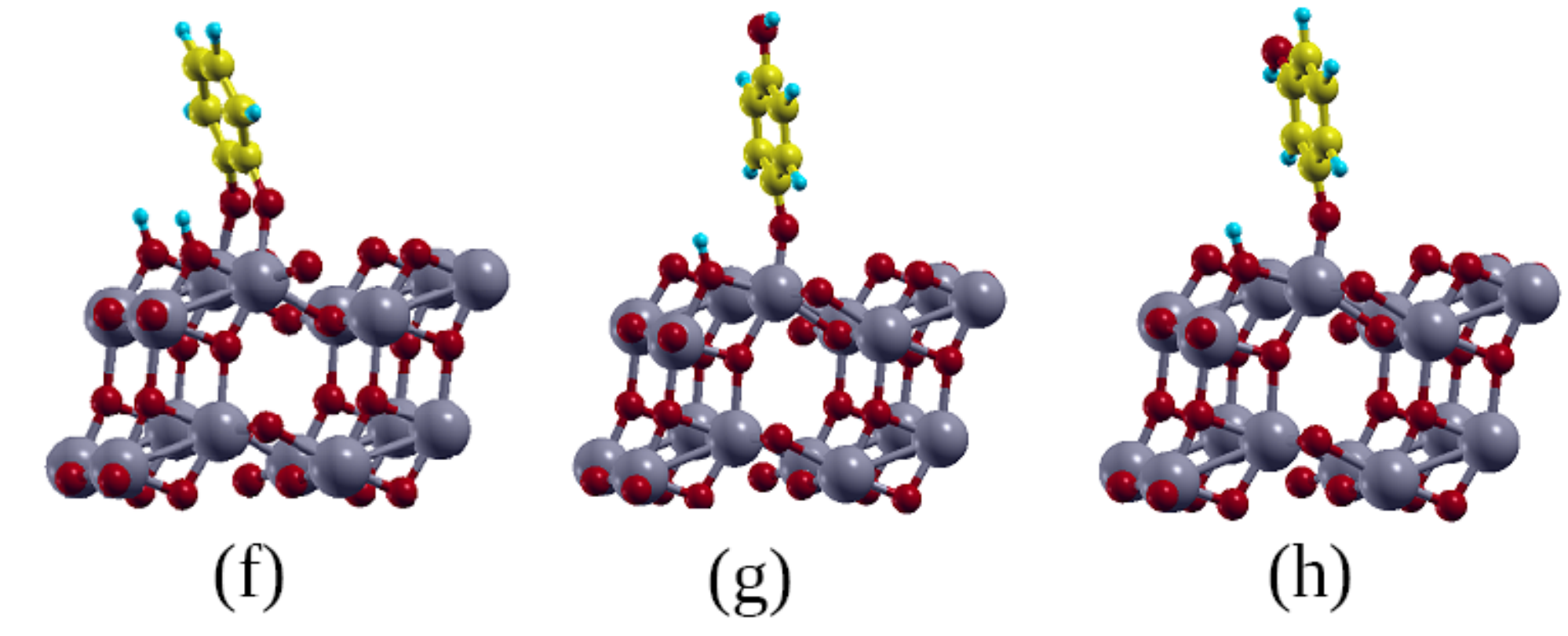}}
\caption{(Color online) Panel (f) shows the bidentate adsorption of catechol on top (101) facet of TiO$_2$ anatase nanowire with double deprotonation, while (g) and (h) represent, respectively, the adsorption geometry of hydroquinone and resorcinol on top (101) facet of the anatase nanowire. }
\label{fig-smp2}
\end{figure}

As precursors, we study dihydroxybenzene isomers group consisting of catechol, hydroquinone and resorcinol. The chemical formula of those isomers is C$_{6}$H$_{4}$(OH)$_{2}$ in which two hydroxyl groups are substituted onto a benzene ring in ortho, para and meta positions, respectively [panels (c), (d) and (e) of figure~\ref{fig-smp1}]. The presence of anchoring groups in those molecules makes them capable of adsorbing dissociatively on the surface of TiO$_2$ nanowire. For catechol, it has three possible ways to adsorb on a TiO$_2$ surface  \cite{Tho12}. In a monodentate structure, only one of the oxygen atoms is bonded to a titanium one. In a bridging bidentate structure, each oxygen is bonded to a different titanium atom on the surface. In the bidentate chelating structure, both oxygen atoms are bonded to the same titanium atom. However, Redfern et al. \cite{Red03} have shown, in a theoretical calculations that a catechol molecule should adsorb on the anatase (101) surface in a bidentate bridging structure. Hence, we focus here on the bridging bidentate adsorption geometry of catechol on the nanowire, which corresponds to the most stable configuration [panel (f) of figure~\ref{fig-smp2}]. For hydroquinone and resorcinol, their design makes them capable of adsorbing on the nanowire surface through one hydroxyl group only [panels (g), (h) of figure~\ref{fig-smp2}].

%%%
\section{Results and discussion }
\subsection{Dihydroxybenzenes adsorption on titanium dioxide and their effect on  {\it  E. Coli} inactivation } \label{uc-part}
In dark conditions, the addition of titanium dioxide to a solution containing one of the three C$_{6}$H$_{4}$(OH)$_{2}$ isomers decreases the concentration of this latter. Moreover, for 4 hours,  C$_{6}$H$_{4}$(OH)$_{2}$ isomer which has the highest concentration decrease is catechol, followed by resorcinol, and finally hydroquinone \cite{Rin01}. In order to understand this experimental result, we calculate the adsorption energy ($E_{\text{ads}}$) of each structure of figure~\ref{fig-smp2}. This parameter, which represents the adsorption degree of each  C$_{6}$H$_{4}$(OH)$_{2}$ isomer on TiO$_2$ nanowire, is calculated according to the following equation:
\begin{equation}
\label{Adsorption energy}
E_{\text{ads}}=E_{\text{wire}}+E_{\text{molecule}}-E_{\text{wire+molecule}}.
\end{equation}

By analyzing the values of adsorption energies gathered in table~\ref{tbl-smp1}, we can conclude that catechol has the highest adsorption energy, followed by resorcinol, then hydroquinone. Consequently, according to the sequence of concentration decrease of those chemical substances, in the presence of TiO$_2$, in the dark, we can conclude that this order is due to the chemical adsorption degree. The C$_{6}$H$_{4}$(OH)$_{2}$ that has the highest concentration decay (catechol), is the most adsorbed on TiO$_2$ surface. In fact, in the case of catechol, the formation of the complex with the TiO$_2$ surface is due to the ortho-position hydroxyl groups in its structure. Thus, it has two sites of fixation to the catalyst, in contrast to the case of the other two isomeric dihydroxybenzenes, resorcinol in meta and hydroquinone in para position.

%%%%%%%%%
\begin{table}[!h]
\caption{Table of adsorption energies ($E_{\text{ads}}$) of catechol, resorcinol and  hydroquinone on the anatase nanowire. The adsorption energy is positive according to the convention that we have considered and it increases with the stability of the structure. }
\label{tbl-smp1}
\vspace{2ex}
\begin{center}
\begin{tabular}{|c|c|c|c|c|c|}
\hline\hline
Molecule & $E_{\text{wire+molecule}}$ (eV) & $E_{\text{wire}}$ (eV) & $E_{\text{molecule}}$ (eV)& $E_{\text{ads}}$ (eV)&  $E_{\text{ads}}$ (Kcal/mol) \\
\hline
\hline
Catechol & $- 41332.601$   &  $- 39434.037$   & $- 1897.705$  &  0.859  & 19.809\\
\hline
Resorcinol & $- 41329.348$ &  $- 39434.037$    &  $- 1895.060$  &  0.251 & 5.788 \\
\hline
Hydroquinone & $- 41329.096$ &   $- 39434.037$   &  $- 1895.031$   & 0.028 &  0.645 \\
\hline\hline
\end{tabular}
\end{center}
\end{table}

The interaction between TiO$_2$ and dihydroxybenzenes was also studied experimentally by Rinc{\'o}n and Pulgarin \cite{Rin04} in the absence of {\it E.~Coli} in dark conditions. They show that the three isomers have an adsorption capacity in the dark. The same experiment provides the value of the equilibrium constant for each dihydroxybenzene adsorption and shows that the adsorption property decreases from catechol to hydroquinone, which is consistent with our calculations stated in table~\ref{tbl-smp1}.

In dark conditions, dihydroxybenzenes inactivate {\it E.~Coli} both in the presence and in the absence of titanium dioxide. The addition of relatively high concentration of one dihydroxybenzene inactivates {\it E.~Coli} bacteria due to a specific toxic effect on {\it E.~Coli}. The effect of C$_{6}$H$_{4}$(OH)$_{2}$ on {\it E.~Coli} inactivation increases, respectively, from resorcinol to hydroquinone. However, the addition of TiO$_2$ inactivates the corresponding bacteria within the following sequence: hydroquinone $\rightarrow$ resorcinol $\rightarrow$ catechol \cite{Rin01}.

By analyzing the change in {\it E.~Coli} inactivation rate, after TiO$_2$ addition in dark conditions, and according to our results gathered in table~\ref{tbl-smp1}, we can assume that, adsorption degree of  C$_{6}$H$_{4}$(OH)$_{2}$ on TiO$_2$ pronouncedly influences {\it E.~Coli} inactivation. The C$_{6}$H$_{4}$(OH)$_{2}$ substance that is more adsorbed on TiO$_2$ surface (catechol) protects the bacteria from its own bactericidal effect which is blocked by adsorption on TiO$_2$. For TiO$_2$, it should be noted that it does not have any effect on {\it E.~Coli} inactivation in dark conditions \cite{Rin01,Mat95}.

It has also been proposed that, in the presence of TiO$_2$, bacterial survival showed to be inversely related to the decay of dihydroxybenzenes concentration due to the adsorption and bacterial intake \cite{Rin04} since, in the dark, the adsorption of dihydroxybenzenes on TiO$_2$ is found to be the most important interaction which limits their action toward the bacteria.
\subsection{Optical properties of C$_{6}$H$_{4}$(OH)$_{2}$ isomers and their effect on {\it E.~Coli} inactivation}\label{ud-part}
\begin{figure}[!t]
\centerline{\includegraphics[width=0.65\textwidth]{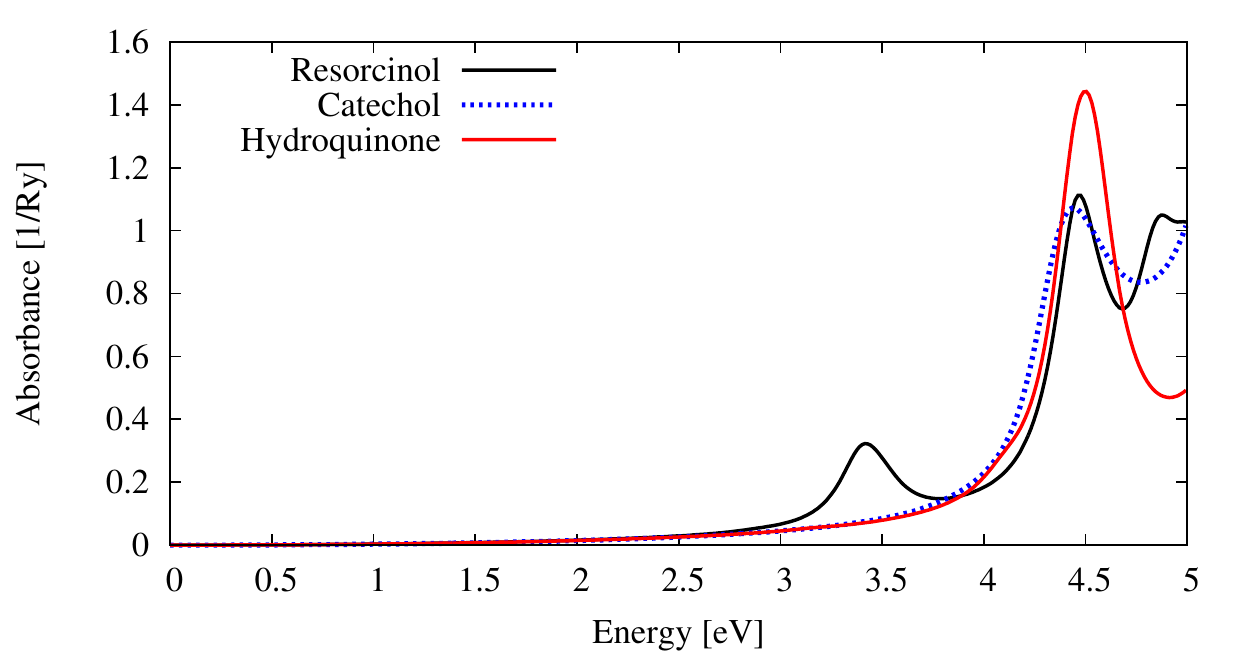}}
\caption{(Color online) TDDFT optical spectra of catechol, hydroquinone and resorcinol.}
\label{fig-smp3}
\end{figure}

Under sunlight effect, the total time of {\it E.~Coli} inactivation is found to be shorter in the absence of dihydroxybenzenes rather than in their presence \cite{Rin01}. Hence, in order to understand the negative effect of dihydroxybenzenes on {\it E.~Coli} inactivation during illumination, we have calculated the optical absorption spectra of all C$_{6}$H$_{4}$(OH)$_{2}$ isomers (figure~\ref{fig-smp3}). We find that all the three dihydroxybenzenes absorb in UV range. The lowest absorption energy of resorcinol is 3.42~eV. This value is consistent with the experiment which shows that resorcinol-based ultraviolet absorbers are highly effective in filtering harmful UV-A and UV-B rays over a broad spectrum [$\lambda_{\text{max}}$ from about 280 to about 400~nm (that is from 3.1 to 4.42~eV)] \cite{Mas03}. For catechol, our calculations show that this energy (the lowest absorption energy) equals 4.44~eV, which is very close to the experimental value, i.e., 4.51~eV \cite{Dun05}. Concerning hydroquinone, it should be noted that its lowest absorption energy equals 4.49~eV. For the experimental absorption band of this latter substance, Pedro Lopez Garcia et al. have shown that hydroquinone absorbs photons belonging to the UV region between 190 and 350~nm that is between 3.54 and 6.52~eV \cite{Gar07}.

For the effect of sunlight on micro-organisms, it is confirmed \cite{Pul04} that the synergistic effect of the UV and heating of water by infrared radiation are capable of inactivating the bacteria, but optical effects on this latter are predominant in comparison to the thermal ones. The process of the damage caused by UV light on {\it E.~Coli} was widely detailed in a recent study \cite{Mur17}.

The ability of dihydroxybenzenes to absorb UV irradiation makes longer the total time needed for bacterial inactivation, when adding those substances. Their presence protects {\it E.~Coli} from a part of UV photons which would affect these bacteria.  For catechol and hydroquinone, they absorb photons belonging to ultraviolet~C (4.43--12.4~eV), which is germicidal. For resorcinol, it absorbs photons of ultraviolet~A (3.10--3.94~eV) and ultraviolet~B (3.93--4.42~eV) which have longer wavelengths. Thereby, catechol and hydroquinone protect the bacteria from the most energetic photons compared to resorcinol. Therefore, resorcinol, which absorbs the light less, is also the substance that less protects the bacteria, and consequently, the total time of bacterial abatement is shorter in the presence of resorcinol in comparison with its other isomers.

For UV-C irradiation, it should be noted that this light is usually absorbed by the atmosphere without reaching the earth surface. Most of papers  which study the effect of UV-C range on micro-organisms utilize artificial UV-C. By comparing the effect of UV-A and C on the bacteria, Paleologou et al. \cite{Pal07}, proved that disinfection by UV-C is substantially effective yielding up to 100\% inactivation with no bacterial regrowth. However, after the same contact time, UV-A irradiation does not cause damage to all the existing bacteria.

In the same context, Pigeot-Remy et al. \cite{Pig12} had studied the effect of UV-A, UV-B and UV-C on {\it E.~Coli} cells. They had proven that the photolysis mechanism of UV-A and UV-B radiations on cell inactivation is dissimilar to the mechanism of UV-C photons, but, the latter radiation is found to be the most efficient treatment to induce a rapid loss of cultivability of {\it E.~Coli} bacterial cells.

\section{Conclusion}
Having calculated the adsorption energy of C$_{6}$H$_{4}$(OH)$_{2}$ isomers on titanium dioxide using DFT, we have found that catechol has the highest adsorption degree, followed by resorcinol, and finally hydroquinone which explains the order of concentration's decay of each one of those isomers in the presence of TiO$_2$. The C$_{6}$H$_{4}$(OH)$_{2}$ that has the lowest concentration decay (hydroquinone), is the less adsorbed on TiO$_2$ surface.

Moreover, in dark conditions, the comparison between the effect of dihydroxybenzenes on {\it E.~Coli} in the absence and in the presence of TiO$_2$ confirms that the adsorption of dihydroxybenzene isomers on titanium dioxide limits the action of those C$_{6}$H$_{4}$(OH)$_{2}$ substances on {\it E.~Coli} inactivation.
Concerning the optical properties of the three C$_{6}$H$_{4}$(OH)$_{2}$ isomers, our calculations confirm that all of dihydroxybenzenes absorb UV irradiation. However, the fact that resorcinol absorbs less energetic photons in comparison to hydroquinone and catechol explains its lower protection to {\it E.~Coli} than its two other isomers.
\section*{Acknowledgements}
The authors would like to acknowledge financial support by the ICTP through Federation scheme. We also thank the CNRST/MaGrid providing the technical support, computing and storage facilities.

\ukrainianpart
\title{Вплив  C$_{6}$H$_{4}$(OH)$_{2}$ ізомерів  на дезинфекцію води через фотокаталіз: чисельні розрахунки}
\author{K. Елменауар, Р. Бенбрік, A. Аамуше}
\address{Дослідницька група MSISM, фізичний відділ, полідисциплінарний факультет Сафі, університет Каді Аяда,  46000, Сафі, Марокко }

\makeukrtitle
\begin{abstract}
Сонячна дезинфекція шляхом фотокаталізу є одним з обіцяючих методів, що використовується для очистки питної води. Вона веде до  руйнування таких бактерій як \textit{Escherichia Coli} (\textit{E.~Coli}).  В цій статті ми порівнюємо наші теоретичні результати з експериментальними, які отримав раніше А.Г. Рінкон і його колеги стосовно ступеня розпаду C$_{6}$H$_{4}$(OH)$_{2}$ ізомерів в присутності діоксиду титану  TiO$_{2}$ і щодо
 впливу оптичних властивостей  тих молекул на інактивацію {\it E.~Coli}. Згідно параметра енергії адсорбції, ми знайшли, що катехол має найвищу ступінь адсорбції на діоксиді титану, після якого іде резорцин і, накінець, гідрохінон.
   Три з ізомерів дигідроксибензолу абсорбують фотони, які належать до ультрафіолетового діапазону. Найнижчі енергії абсорбції  резорцину, катехолу і гідрохінону є, відповідно, 3.42, 4.44 and 4.49~еВ.
\keywords  фотокаталіз, {\it E.~Coli}, TiO$_2$, C$_{6}$H$_{4}$(OH)$_{2}$, спектр абсорбції, квантове ESPRESSO

\end{abstract}

\begin{thebibliography}{99}

\bibitem{Coo93} Cooper W.J., Cadavid E., Nickelsen M.G., Lin  K.J., Kurucz C.N., Waite T.D., J. Am. Water Works Assn., 1993, \textbf{85}, 106.
\bibitem{Ric96} Richardson  S.D., Thruston A.D. (Jr.), Collette T.W. , Patterson  K.S., Lykins B.W. (Jr.), Ireland J.C., Environ. Sci. Technol., 1996, \textbf {30}, 3327, \doi{10.1021/es960142m}.
\bibitem{Par12} Paramasivam I., Jha H., Liu N., Schmuki P., Small, 2012, \textbf{8}, 3073, \doi{10.1002/smll.201200564}.
\bibitem{Sch14} Schneider J., Matsuoka M., Takeuchi M., Zhang J., Horiuchi Y., Anpo M., Bahnemann D.W.,  Chem. Rev., 2014, \textbf{114}, 9919, \doi{10.1021/cr5001892}.
\bibitem{Sas10} Sasahara A., Onishi H., Solid State Phenom., 2010, \textbf{162}, 115, \doi{10.4028/www.scientific.net/SSP.162.115}.
\bibitem{Hoa12} Hoang S., Berglund S.P., Hahn N.T., Bard A.J., Mullins C.B., J. Am. Chem. Soc., 2012, \textbf{134}, 3659, \\ \doi{10.1021/ja211369s}.
\bibitem{Lee14} Lee K., Mazare A., Schmuki P., Chem. Rev., 2014, \textbf{114}, 9385, \doi{10.1021/cr500061m}.
\bibitem{Hab13} Habisreutinger S.N., Schmidt-Mende L., Stolarczyk J.K., Angew. Chem. Int. Ed., 2013, \textbf{52}, 7372,\\ \doi{10.1002/anie.201207199}.
\bibitem{Roy11} Roy P., Berger S., Schmuki P., Angew. Chem. Int. Ed., 2011, \textbf{50}, 2904, \doi{10.1002/anie.201001374}.
\bibitem{Cha08} Chanmanee W., Watcharenwong A., Chenthamarakshan C.R., Kajitvichyanukul P., de Tacconi  N.R., Rajeshwar~K., J. Am. Chem. Soc., 2008, \textbf{130}, 965, \doi{10.1021/ja076092a}.
\bibitem{Li15} Li X., Yu J., Low J., Fang Y., Xiao J., Chen X., J. Mater. Chem. A, 2015, \textbf{3}, 2485, \doi{10.1039/c4ta04461d}.
\bibitem{Tan15} Moniz S.J.A., Shevlin S.A., Martin D.J., Guo Z.-X., Tang J., Energy Environ. Sci., 2015, \textbf{8}, 731, \doi{10.1039/C4EE03271C}.
\bibitem{Laz01} Lazzeri M., Vittadini A., Selloni A., Phys. Rev. B, 2001, \textbf{63}, 155409, \doi{10.1103/PhysRevB.63.155409}.
\bibitem{Mil98} Milligan P.W., H\"aggblom M.M., Environ. Toxicol. Chem., 1998, \textbf{17}, 1456, \doi{10.1002/etc.5620170804}.
\bibitem{Hoh64} Hohenberg P., Kohn W., Phys. Rev., 1964, \textbf{136}, B864, \doi{10.1103/PhysRev.136.B864}.
\bibitem{Run84} Runge E., Gross E.K.U., Phys. Rev. Lett., 1984, \textbf{ 52}, 997, \doi{10.1103/PhysRevLett.52.997}.
\bibitem{Gia09} Giannozzi P., Baroni S., Bonini N., Calandra  M., Car R., Cavazzoni C., Ceresoli D., Chiarotti G.L.,  Cococcioni~M., Dabo  I., \textit{et al.},
%Corso  A.D., de Gironcoli S., Fabris  S., Fratesi G., Gebauer R., Gerstmann U., Gougoussis~C., Kokalj~A., Lazzeri M., Martin-Samos L., Marzari N., Mauri F., Mazzarello R., Paolini S., Pasquarello A., Paulatto~L., Sbraccia C., Scandolo S., Sclauzero G., Seitsonen A.P., Smogunov A., Umari P., Wentzcovitch~R.M.,
J. Phys.: Condens. Matter, 2009, \textbf{21}, 395502, \doi{10.1088/0953-8984/21/39/395502}.
\bibitem{url1} URL~\url{http://www.quantum-espresso.org}.
\bibitem{Per96} Perdew J.P., Burke K., Ernzerhof M., Phys. Rev. Lett., 1996, 77, 3865, \doi{10.1103/PhysRevLett.77.3865}.
\bibitem{Wal06} Walker B., Saitta A.M., Gebauer R., Baroni S., Phys. Rev. Lett., 2006, \textbf{96}, 113001,\\ \doi{10.1103/PhysRevLett.96.113001}.
\bibitem{Roc08} Rocca D., Gebauer R., Saad Y., Baroni S., J. Chem. Phys., 2008, \textbf{128}, 154105, \doi{10.1063/1.2899649}.
\bibitem{Bar10} Baroni S., Gebauer R., Malcio{\u g}lu O.B., Saad Y., Umari P., Xian J., J. Phys.: Condens. Matter, 2010, \textbf{22}, 074204, \doi{10.1088/0953-8984/22/7/074204}.
\bibitem{Wal08} Walker B.G., Hendy S.C., Gebauer R., Tilley R.D., Eur. Phys. J. B, 2008, \textbf{66}, 7, \doi{10.1140/epjb/e2008-00388-1}.
\bibitem{Gho10} Ghosh P., Gebauer R., J. Chem. Phys., 2010, \textbf{132}, 104102, \doi{10.1063/1.3326226}.
\bibitem{Mal11} Malcio{\u g}lu O.B., Gebauer R., Rocca D., Baroni S., Comput. Phys. Commun., 2011, \textbf{182}, 1744,\\ \doi{10.1016/j.cpc.2011.04.020}.
\bibitem{Tho12} Thomas A.G., Syres K.L., Chem. Soc. Rev., 2012, \textbf{41}, 4207, \doi{10.1039/C2CS35057B}.
\bibitem{Red03} Redfern P.C., Zapol P., Curtiss L.A., Rajh T., Thurnauer M.C., J. Phys. Chem. B, 2003, \textbf{107}, 11419, \doi{10.1021/jp0303669}.
\bibitem{Rin01} Rinc{\'o}n A.G., Pulgarin C., Adler N., Peringer P., J. Photochem. Photobiol. A, 2001, \textbf{139}, 233,\\ \doi{10.1016/S1010-6030(01)00374-4}.
\bibitem{Rin04} Rinc{\'o}n A.G., Pulgarin C., Appl. Catal. B, 2004, \textbf{51}, 283, \doi{10.1016/j.apcatb.2004.03.007}.
\bibitem{Mat95} Matsunaga T., Okochi M., Environ. Sci. Technol., 1995, \textbf{29}, 501, \doi{10.1021/es00002a028}.
\bibitem{Mas03} Mason M., Zhao X., Stephen E., US Patent App. 09/934, 374, 2003, \textbf{US 2003/0078328 A1}.
\bibitem{Dun05} Duncan W.R., Prezhdo O.V., J. Phys. Chem. B, 2005, \textbf{109}, 365, \doi{10.1021/jp046342z}.
\bibitem{Gar07} Garcia P.L., Santoro M.I.R.M., Singh A.K., Kedor-Hackmann E.R.M., Braz. J. Pharm. Sci., 2007, \textbf{43}, 397.
\bibitem{Pul04} Rinc{\'o}n A.G., Pulgarin C., Sol. Energy, 2004, \textbf{77}, 635, \doi{10.1016/j.solener.2004.08.002}.
\bibitem{Mur17} Murcia J.J., {\'A}vila-Mart\'inez E.G., Rojas H., Nav{\'i}o J.A., Hidalgo M.C., Appl. Catal. B, 2017, \textbf{200}, 469, \doi{10.1016/j.apcatb.2016.07.045}.
\bibitem{Pal07} Paleologou A., Marakas H., Xekoukoulotakis N.P., Moya A., Vergara Y., Kalogerakis N., Gikas P., Mantzavinos~D., Catal. Today, 2007, \textbf{129}, 136, \doi{10.1016/j.cattod.2007.06.059}.
\bibitem{Pig12} Pigeot-R\'emy S., Simonet F., Atlan D., Lazzaroni J.C., Guillard C., Water Res., 2012, \textbf{46}, 3208,\\ \doi{10.1016/j.watres.2012.03.019}.
\end{thebibliography}
\end{document}